\documentclass[twocolumn,showpacs,superscriptaddress,amsmath,amssymb,aps,pra]{revtex4-1}
\usepackage{graphicx}
\usepackage{dcolumn}
\usepackage{amsmath,bm}
\usepackage{epstopdf}
\usepackage[mathlines]{lineno}
\usepackage{color}
\usepackage[colorlinks=true,linkcolor=blue,citecolor=blue]{hyperref}

\begin{document}
\title{Anisotropic exchange coupling in a nanowire double quantum dot with strong spin-orbit coupling}
\author{Rui\! Li}
\email{rl.rueili@gmail.com}
\affiliation{Beijing Computational Science Research Center, Beijing 100084, China}

\author{J.\!\! Q.\! You}
\email{jqyou@csrc.ac.cn}
\affiliation{Beijing Computational Science Research Center, Beijing 100084, China}
\affiliation{Synergetic Innovation Center of Quantum Information and Quantum Physics, University of Science and Technology of China, Hefei, Anhui 230026, China}

\begin{abstract}
A spin-orbit qubit is a hybrid qubit that contains both orbital and spin degrees of freedom of an electron in a quantum dot. Here we study the exchange coupling between two spin-orbit qubits in a nanowire double quantum dot (DQD) with strong spin-orbit coupling (SOC). We find that while the total tunneling in the DQD is irrelevant to the SOC, both the spin-conserved and spin-flipped tunnelings are SOC dependent and can compete with each other in the strong SOC regime. Moreover, the Coulomb repulsion between electrons can combine with the SOC-dependent tunnelings to yield an anisotropic exchange coupling between the two spin-orbit qubits. Also, we give an explicit physical mechanism for this anisotropic exchange coupling.
\end{abstract}

\date{\today}
\maketitle

\section{Introduction}
Realizing a controllable interqubit coupling is of essential importance in quantum information processing (see, e.g., Refs.~\cite{Ladd,Xiang}). For the electron spin qubit defined in a semiconductor quantum dot~\cite{Hanson}, the two-qubit coupling can be achieved using the isotropic Heisenberg exchange interaction in a tunneling-coupled double quantum dot (DQD)~\cite{Burkard,Xuedong}. Recently, a hybrid qubit, the spin-orbit qubit~\cite{Nadj1,Nadj2}, was achieved in a nanowire quantum dot with strong spin-orbit coupling (SOC). A distinct advantage of this spin-orbit qubit is its manipulability via an electric field (an effect called electric-dipole spin resonance~\cite{Nadj1,Nadj2,RuiLi,Nowack,Pioro,Laird,Rashba,Tokura,Golovach,Sherman}) because a local electric field can be generated in experiments much more easily than a local magnetic field~\cite{Koppens}.

The key element for achieving a spin-orbit qubit is the availability of strong SOC in a quantum-dot material. The semiconductor nanowire materials, e.g., InAs~\cite{Nadj1,Takahashi,Schroer,Petersson} and InSb nanowire~\cite{Nadj2,Pribiag}, provide an ideal platform for realizing such a qubit. Indeed, a large Rabi frequency of~$\sim$100 MHz was reported recently for single-qubit operations~\cite{van}. Interestingly, in the presence of strong SOC, the coupling between the spin-orbit qubit and the electric field depends nonlinearly on the SOC strength~\cite{Trif} and there is an optimal SOC where the Rabi frequency induced by an ac electric field becomes maximal~\cite{RuiLi}. Now, it becomes desirable to realize a controllable coupling between two spin-orbit qubits, in order to implement nontrivial (i.e., conditional) two-qubit operations.

In this paper, we investigate the exchange coupling between two spin-orbit qubits in a gated semiconductor nanowire DQD with strong SOC. Our main goal is to clarify the effect of the strong SOC on the exchange coupling. First, we derive a second quantized Hamiltonian for the DQD, where the electron field operator is expanded in terms of the spin-orbit basis~\cite{RuiLi}, other than the conventional basis with separable spin and orbital degrees of freedom~\cite{Baruffa} which is valid only in the zero or weak SOC regime. We find that there exist both spin-conserved tunneling $t$ and spin-flipped tunneling $t'$ in the DQD, where the mentioned spin is actually a pseudo-spin (spin-orbit qubit)~\cite{RuiLi}. It is interesting to note that $|t|^{2}+|t'|^{2}$ is irrelevant to the SOC, but $t'$ can compete with $t$ when increasing the SOC. Then, we study the exchange coupling by considering two electrons confined in this nanowire DQD. In the strong SOC regime, our results reveal that in contrast to the usual isotropic exchange coupling, the Coulomb repulsion between electrons can combine with the SOC-dependent tunnelings $t$ and $t'$ to yield an anisotropic exchange coupling between the two spin-orbit qubits. We explicitly explain the physical mechanism of this anisotropic exchange coupling and show that the obtained energy spectrum of the two coupled spin-orbit qubits is qualitatively in good agreement with the recent experimental results.

The paper is organized as follows. In Sec.~\ref{secII}, we give analytical expressions for the SOC-dependent tunnelings in a nanowire DQD which are valid in the strong SOC regime. In Sec.~\ref{secIII}, we study the exchange coupling between two spin-orbit qubits in this nanowire DQD. Also, the impacts of the strong SOC are explicitly clarified. Finally, we conclude in Sec.~\ref{secIV}.

\section{\label{secII}SOC-dependent tunneling in a nanowire DQD}
\begin{figure}
\includegraphics{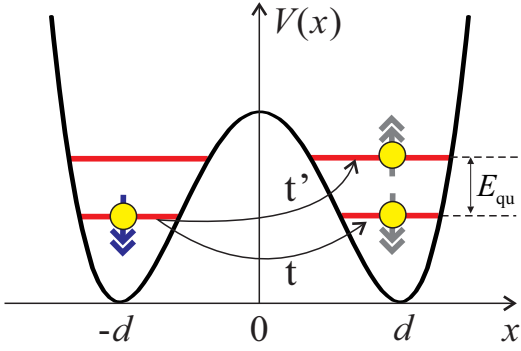}
\caption{\label{Fig_Double_quantum_dot}A nanowire DQD modeled by a double-well potential. Both spin-conserved tunneling $t$ and spin-flipped tunnelings $t'$ exist in the DQD due to the presence of the SOC. $E_{\rm qu}=E_{l(r)\Uparrow}-E_{l(r)\Downarrow}$ is the level spacing of the spin-orbit qubit and the double arrows represent the basis states of the spin-orbit qubit (i.e., the pseudospin).}
\end{figure}

It is interesting to first clarify the effects of the strong SOC on the electron tunneling in a DQD, because previous studies only focused on the weak SOC regime~\cite{Moriya,Mireles}. Figure~\ref{Fig_Double_quantum_dot} schematically shows the considered semiconductor nanowire DQD with strong SOC, where an electron is confined in a double well and subjected to an external static magnetic field~\cite{Flindt,Khomitsky,Nowak}. The Hamiltonian reads
\begin{equation}
H=p^{2}/(2m_{e})+V(x)-\alpha\sigma^{y}p+(g_{e}\mu_{B}B/2)\sigma^{x},\label{Eq_DQD_Hamiltonian}
\end{equation}
where $p=-i\hbar\partial/\partial\,x$, $V(x)$ is the double-well potential characterizing the DQD, $\alpha$ is the Rashba SOC strength~\cite{bychkov2}, and an external static magnetic field $\textbf{B}$ is applied in the $x$ direction. For simplicity, we consider a symmetric double-well potential (see Fig.~\ref{Fig_Double_quantum_dot}).

In order to explicitly show the role that the strong SOC plays in a DQD, we need to derive a second quantized Hamiltonian for the DQD. Similarly to the derivation of the tight-binding Hamiltonian, we first calculate the localized wave function centered at each dot and then expand the electron field operator in terms of these localized wave functions.

Near the minimum of each well, the potential can be expanded harmonically as $V(x)=\frac{1}{2}m_{e}\omega^{2}(x\pm\,d)^{2}+\cdots$, with $2d$ being the interdot distance. Thus, we have the following Hamiltonian which describes an electron localized in either dot:
\begin{equation}
H_{l/r}=\frac{p^{2}}{2m_{e}}+\frac{1}{2}m_{e}\omega^{2}(x\pm\,d)^{2}-\alpha\sigma^{y}p+\frac{g_{e}\mu_{\rm B}B}{2}\sigma^{x}.\label{Eq_single_QD}
\end{equation}
In order to capture all the information of the SOC, we only treat the Zeeman term, instead of the SOC, as perturbation~\cite{Levitov,RuiLi}, which is valid when $g_{e}\mu_{B}B/(\hbar\omega)\ll1$. As estimated in Ref.~\cite{RuiLi} for an InSb nanowire quantum dot, the external static magnetic field can be as strong as $B\sim0.1$ T, which is larger than the magnetic field usually used in a quantum device. The lowest two states of Eq.~(\ref{Eq_single_QD}), up to zeroth order, are
\begin{equation}
\phi_{l\sigma}(x)=\phi_{\sigma}(x+d),~~~\phi_{r\sigma}(x)=\phi_{\sigma}(x-d),\label{EQ_unorthonormal}
\end{equation}
where $\sigma=$~$\Uparrow$ and $\Downarrow$ describe the two pseudospin states, and the wave functions are given by~\cite{Rashba2}
\begin{eqnarray}
\phi_{\Uparrow}(x)&=&\psi_{0}(x)\left[\cos(x/x_{\rm so})|\!\uparrow_{x}\rangle-\sin(x/x_{\rm so})|\!\downarrow_{x}\rangle\right],\nonumber\\
\phi_{\Downarrow}(x)&=&-i\psi_{0}(x)\left[\cos(x/x_{\rm so})|\!\downarrow_{x}\rangle+\sin(x/x_{\rm so})|\!\uparrow_{x}\rangle\right].~~~\label{EQ_spin_orbit_basis}
\end{eqnarray}
Here $x_{\rm so}=\hbar/(m_{e}\alpha)$ is the spin-orbit length, $|\!\!\!\uparrow_{x}\rangle$ and $|\!\!\!\downarrow_{x}\rangle$ are the eigenstates of $\sigma^{x}$, and $\psi_{0}(x)=\left[m_{e}\omega/(\hbar\pi)\right]^{1/4}{\rm exp}[-x^{2}/(2x^{2}_{0})]$ is the ground state of the harmonic oscillator. The corresponding eigenvalues of $\phi_{\Uparrow/\Downarrow}$ are
\begin{eqnarray}
E_{l\Uparrow/\Downarrow}=E_{r\Uparrow/\Downarrow}&=&(1/2)\hbar\omega\left[1-(x_{0}/x_{\rm so})^{2}\right]\nonumber\\
&&\pm\,g_{e}\mu_{B}B{\rm exp}[-(x_{0}/x_{\rm so})^{2}],
\end{eqnarray}
where $x_{0}=\sqrt{\hbar/(m_{e}\omega)}$ defines a characteristic length.

Note that these four states are not orthogonal, because there are overlap integrations among them:
\begin{eqnarray}
s_{a}&=&\int\,dx\phi^{\dagger}_{r\sigma}(x)\phi_{l\sigma}(x)={\rm exp}(-d^{2}/x^{2}_{0})\cos(2d/x_{\rm so}),\nonumber\\
s_{b}&=&\int\,dx\phi^{\dagger}_{r\sigma}(x)\phi_{l\bar{\sigma}}(x)=-i{\rm exp}(-d^{2}/x^{2}_{0})\sin(2d/x_{\rm so}).~~~~\label{Eq_overlapintegration}
\end{eqnarray}
It can be seen that due to the SOC, the overlap integration $s_{b}$ becomes nonzero. This is different from the case of zero SOC~\cite{Burkard}, where $x_{\rm so}\rightarrow\infty$, so $s_{a}={\rm exp}(-d^{2}/x^{2}_{0})$ and $s_{b}=0$. Based on these four localized wave functions,
we can derive an orthonormal basis $\phi^{\rm or}_{k\sigma}(x)$ via the Schmidt orthogonalization (for details see Appendix~\ref{appendix_A}).

The electron field operator can be expanded in terms of the orthonormal basis $\Psi(x)=\sum_{k=l,r;\sigma=\Uparrow,\Downarrow}c_{k\sigma}\phi^{\rm or}_{k\sigma}(x)$, where $\phi^{\rm or}_{k\sigma}(x)$ form the spin-orbit basis, in which both the spin and the orbital states are entangled due to the SOC. This is in sharp contrast to the usual basis where both the spin and the orbital states constitute a product state~\cite{Baruffa,Moriya,Kavokin}. In the presence of the strong SOC, the electron spin is no longer conserved in the quantum dot. Therefore, when an electron is injected into the quantum dot, the electron should occupy spin-orbit basis states (i.e., the eigenstates of each dot) instead of the product basis states of the spin and the orbit. It should be noted that other excited orbits are not considered here, because they are well separated from the lowest two orbits roughly by $\hbar\omega$, i.e., $E_{l(r)\Uparrow}-E_{l(r)\Downarrow}\ll\hbar\omega$. The DQD Hamiltonian can be calculated as
\begin{eqnarray}
H_{\rm DQD}&=&\int\,dx\Psi^{\dagger}(x)H\Psi(x)=\sum_{\sigma=\Uparrow,\Downarrow}\big[\varepsilon_{l\sigma}c^{\dagger}_{l\sigma}c_{l\sigma}\nonumber\\
&&+\varepsilon_{r\sigma}c^{\dagger}_{r\sigma}c_{r\sigma}+(tc^{\dagger}_{l\sigma}c_{r\sigma}+t'c^{\dagger}_{l\sigma}c_{r\bar{\sigma}}+{\rm h.c.})\big],~~\label{Eq_Second_Quanzied_Hamiltonian}
\end{eqnarray}
where $t$ is the spin-conserved tunneling amplitude and $t'$ is the spin-flipped tunneling amplitude. In previous weak-SOC theories~\cite{Moriya,Mireles}, the spin-flipped terms also exist but $|t'/t|\ll1$. However, in our strong-SOC theory, both $t$ and $t'$ depend nonlinearly on the SOC strength $\alpha$, and the ratio $|t'/t|$ can be even larger than 1 when increasing the SOC strength $\alpha$ (see below).

When the interdot distance is larger than the characteristic length (i.e., $d>x_{0}$), it follows from Eq.~(\ref{Eq_overlapintegration}) that $|s_{a,b}|\rightarrow0$. Now the parameters of the DQD have the following explicit analytical expressions (accurate to the first order of $|s_{a,b}|$):
\begin{eqnarray}
&&\varepsilon_{l\sigma}=\varepsilon_{r\sigma}=E_{l\sigma}=E_{r\sigma},\nonumber\\
&&t=t_{0}\cos(2d/x_{\rm so}),~~~t'=-it_{0}\sin(2d/x_{\rm so}),
\end{eqnarray}
where $t_{0}$ is the interdot tunneling amplitude in the absence of the SOC; e.g., $t_{0}=-3V_{0}{\rm exp}(-d^{2}/x^{2}_{0})$ for a double-well potential $V(x)=V_{0}[(x/d)^{2}-1]^{2}$. As we have emphasized above, the SOC is not treated as a perturbation in our calculations, so these expressions are valid in the strong and even ultrastrong SOC regimes. In the weak SOC limit with $\alpha\rightarrow0$ (i.e., $x_{\rm so}\rightarrow\infty$), we recover the previous results $t\approx\,t_{0}$ and $t'\approx\,-(2id/x_{\rm so})t_{0}$~\cite{Moriya,Mireles}; i.e., the spin-flipped tunneling is proportional to the SOC strength $\alpha$.
\begin{figure}
\includegraphics{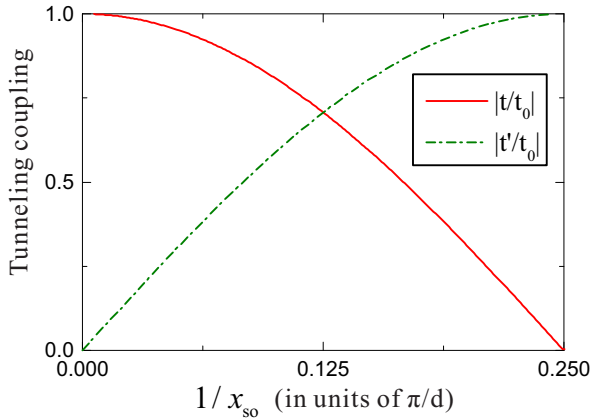}
\caption{\label{Fig_tunneling}The spin-conserved and spin-flipped tunnelings in a DQD as a function of the SOC, where $t_{0}$ corresponds to the tunneling without the SOC.}
\end{figure}

There is something unexpected in the strong SOC regime. As we show in Fig.~{\ref{Fig_tunneling}}, when increasing the SOC, the spin-flipped tunneling $|t'|$ can compete with the spin-conserved tunneling $|t|$, while the total tunneling $|t|^{2}+|t'|^{2}=t^{2}_{0}$ is irrelevant to the SOC. The experimentally measured SOC length in an InSb nanowire is $x_{\rm so}\approx230\pm50$ nm~\cite{Nadj2}. For an InSb DQD with an interdot distance $2d\sim50$ nm, $|t'/t|=\tan(2d/x_{\rm so})\sim0.22$, indicating that the spin-flipped tunneling also becomes appreciable in this device. The interesting competition between $|t|$ and $|t'|$ is owing to the peculiar spin-orbit basis in the strong SOC regime [see Eq.~(\ref{EQ_spin_orbit_basis})].

The SOC can lift the Pauli spin blockade of electron tunneling in a DQD~\cite{Pfund,Fasth,NadjPerge,Schreiber,Danon}. Our result explicitly shows that this reduction is due to the presence of the spin-flipped tunneling. This indicates that the existence of the spin-flipped tunneling can yield important effects on the measurements of a spin-orbit qubit when the DQD is tuned to the Pauli spin blockade regime.

\section{\label{secIII}The anisotropic exchange coupling}
Below we explore how the strong SOC affects the exchange coupling~\cite{Baruffa,Gangadharaiah,Greilich} in the nanowire DQD. It is known that in the absence of the SOC, the spin-orbit qubit is reduced to a spin qubit, and the exchange coupling between two electron spins in a DQD is just the isotropic Heisenberg interaction~\cite{Burkard,Xuedong}.

We consider two electrons confined in a nanowire DQD. The Coulomb interaction between these two electrons is given by
\begin{equation}
H_{U}=\frac{1}{2}\int\,dxdx'\Psi^{\dagger}(x)\Psi^{\dagger}(x')\frac{e^{2}}{|x-x'|}\Psi(x')\Psi(x).
\end{equation}
Including both intra- and interdot Coulomb interactions, we have the Hubbard-like Hamiltonian
\begin{eqnarray}
H&=&\sum_{\sigma=\Uparrow,\Downarrow}\big[\varepsilon_{l\sigma}c^{\dagger}_{l\sigma}c_{l\sigma}+\varepsilon_{r\sigma}c^{\dagger}_{r\sigma}c_{r\sigma}
+(tc^{\dagger}_{l\sigma}c_{r\sigma}+t'c^{\dagger}_{l\sigma}c_{r\bar{\sigma}}\nonumber\\
&&+{\rm h.c.})\big]+Un_{l\Uparrow}n_{l\Downarrow}+Un_{r\Uparrow}n_{r\Downarrow}+U'\sum_{\sigma\sigma'}n_{l\sigma}n_{r\sigma'},~~~~
\end{eqnarray}
where $U$ and $U'$ respectively represent the strengths of the intra- and interdot Coulomb repulsions. Note that $\sigma$ describes the pseudospin states, i.e., the two eigenstates of the spin-orbit qubit. We consider the strong repulsion regime with $(U-U')\gg\,|t|,|t'|$, such that each dot contains only one electron. Thus, we can define a projection operator~\cite{RuiLi2}
\begin{eqnarray}
P&=&[n_{l\Uparrow}(1-n_{l\Downarrow})+n_{l\Downarrow}(1-n_{l\Uparrow})]\nonumber\\
&&\times[n_{r\Uparrow}(1-n_{r\Downarrow})+n_{r\Downarrow}(1-n_{r\Uparrow})],
\end{eqnarray}
which retains the pseudo-spin degrees of freedom of the two electrons but reduces the Hilbert space to the subspace with each dot occupied by one electron. The effective Hamiltonian can be written as\cite{Nagaosa}
\begin{equation}
H_{\rm eff}=PHP-PHQ(QHQ-E)^{-1}QHP,
\end{equation}
where $Q=1-P$. After some algebra, we obtain (for details see Appendix~\ref{appendix_B})
\begin{eqnarray}
H_{\rm eff}&=&E_{\rm qu}(S^{z}_{l}+S^{z}_{r})+(J-J^{[2]}_{\rm so})\textbf{S}_{l}\cdot\textbf{S}_{r}\nonumber\\
&&+J^{[1]}_{\rm so}(\textbf{S}_{l}\times\textbf{S}_{r})_{x}+2J^{[2]}_{\rm so}S^{x}_{l}S^{x}_{r},\label{EQ_anisotropic_exchange}
\end{eqnarray}
where
\begin{eqnarray}
E_{\rm qu}&=&g_{e}\mu_{B}B{\rm exp}[-(x_{0}/x_{\rm so})^{2}],~~J=\frac{4|t|^{2}}{U-U'},\nonumber\\
J^{[1]}_{\rm so}&=&\frac{4i(tt'^{*}-t^{*}t')}{U-U'},~~J^{[2]}_{\rm so}=\frac{4|t'|^{2}}{U-U'},\label{Eq_coefficients}
\end{eqnarray}
and $\textbf{S}_{k=l,r}=(1/2)\sum_{\sigma,\sigma'}c^{\dagger}_{k\sigma}\boldsymbol{\rho}_{\sigma\sigma'}c_{k\sigma'}$ is the pseudospin operator, with $\boldsymbol{\rho}\equiv(\rho^{x},\rho^{y},\rho^{z})$ being the Pauli matrices of the spin-orbit qubit. Therefore, we obtain an anisotropic Heisenberg exchange interaction between the two spin-orbit qubits. The exchange interaction consists of three terms: the antiferromagnetic $J$ term, the anisotropic $J^{[1]}_{\rm so}$ term, and the ferromagnetic $J^{[2]}_{\rm so}$ term. It is known that the SOC introduces an anisotropic exchange $J^{[1]}_{\rm so}$ term in the weak SOC regime~\cite{Moriya,Baruffa,Gangadharaiah}, but the isotropic antiferromagnetic $J$ term dominates. However, in the strong SOC regime, the anisotropic exchange $J^{[1]}_{\rm so}$ term becomes dominant and a ferromagnetic $J^{[2]}_{\rm so}$ term further occurs. This ferromagnetic $J^{[2]}_{\rm so}$ term can even play a dominant role in the ultrastrong coupling regime.

The exchange interaction is induced by the second-order virtual tunneling in a DQD. Each exchange-coupling term in Eq.~(\ref{EQ_anisotropic_exchange}) has an explicit physical picture (for details see Appendix~\ref{appendix_B}): (i) The virtual tunneling involving $t^{2}$ gives an antiferromagnetic exchange interaction $J~\textbf{S}_{l}\cdot\textbf{S}_{r}$, (ii) the virtual tunneling involving the combination of $t$ and $t'$ gives an anisotropic exchange interaction $J^{[1]}_{\rm so}(\textbf{S}_{l}\times\textbf{S}_{r})_{x}$, and (iii) the virtual tunneling involving $t'^{2}$ gives a ferromagnetic exchange interaction $-J^{[2]}_{\rm so}\textbf{S}_{l}\cdot\textbf{S}_{r}+2J^{[2]}_{\rm so}S^{x}_{l}S^{x}_{r}$.

\begin{figure}
\includegraphics[width=8.4cm]{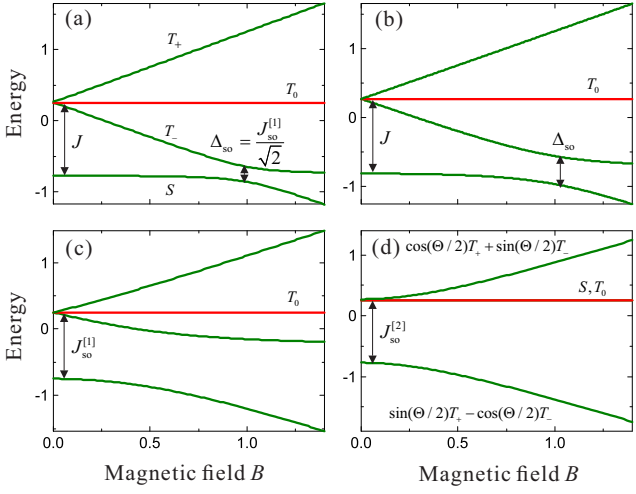}
\caption{\label{Fig_anticrossing}The energy spectrum of two coupled spin-orbit qubits in a DQD with Coulomb repulsion. (a) The spectrum in the weak SOC regime with $|t'/t|\ll1$, where $J\gg\,J^{[1]}_{\rm so}\gg\,J^{[2]}_{\rm so}$. (b) The spectrum calculated using $|t'/t|=0.3$, which is chosen to fit the experimentally measured value $\Delta_{\rm so}/J\approx0.4$ in Ref.~\cite{Nadj2}. (c) The spectrum in the strong SOC regime with $|t'/t|=1$, where $J=J^{[2]}_{\rm so}=(1/2)J^{[1]}_{\rm so}$. (d) The spectrum in the ultrastrong SOC regime with $|t'/t|\gg1$, where $J^{[2]}_{\rm so}\gg\,J^{[1]}_{\rm so}\gg\,J$, and $\Theta=\arctan\big[J^{[2]}_{\rm so}/(2E_{\rm qu})\big]$. In both (a) and (b), the energy is in units of $J$ and the magnetic field $B$ is in units of $J/\mu$ with $\mu=g_{e}\mu_{B}{\rm exp}[-(x_{0}/x_{\rm so})^{2}]$. In (c) and (d), the energy is in units of $J^{[i]}_{\rm so}$ and $B$ is in units of $J^{[i]}_{\rm so}/\mu$, where $i=1$ for (c) and 2 for (d).}
\end{figure}

In the weak SOC regime with $|t'/t|\ll\,1$, $J^{[2]}_{\rm so}\ll\,J^{[1]}_{\rm so}\ll\,J$ in Eq.~(\ref{EQ_anisotropic_exchange}). After neglecting the second-order terms, the effective Hamiltonian (\ref{EQ_anisotropic_exchange}) is reduced to
\begin{equation}
H_{\rm eff}=E_{\rm qu}(S^{z}_{l}+S^{z}_{r})+J~\textbf{S}_{l}\cdot\textbf{S}_{r}
+J^{[1]}_{\rm so}(\textbf{S}_{l}\times\textbf{S}_{r})_{x}.
\end{equation}
The energy spectrum of this Hamiltonian is shown in Fig.~\ref{Fig_anticrossing}(a), where an anticrossing gap (i.e., the spin-orbit gap) $\Delta_{\rm so}=J^{[1]}_{\rm so}/\sqrt{2}$ between singlet state $S$ and triplet state $T_{-}$ occurs due to the anisotropic exchange term $J^{[1]}_{\rm so}(\textbf{S}_{l}\times\textbf{S}_{r})_{x}$. This energy spectrum is qualitatively in good agreement with the experimental results in an InAs nanowire DQD~\cite{Fasth,Greilich}. It is interesting to relate these quantities to the SOC strength
\begin{equation}
x_{\rm so}=2d\times\arctan^{-1}\big[J^{[1]}_{\rm so}/(2J)\big].\label{Eq_SOClength}
\end{equation}
Because both the gap $\Delta_{\rm so}=J^{[1]}_{\rm so}/\sqrt{2}$ at the anticrossing point and the singlet-triplet splitting $J$ are experimentally measurable quantities~\cite{Fasth,Greilich}, one can use Eq.~(\ref{Eq_SOClength}) to obtain the SOC strength $\alpha$ via $x_{\rm so}=\hbar/(m_{e}\alpha)$.

Figure~\ref{Fig_anticrossing}(b) shows the result calculated using Eq.~(\ref{EQ_anisotropic_exchange}) to fit the experimental data in an InSb nanowire DQD with $\Delta_{\rm so}\approx0.4J$~\cite{Nadj2}. In this fitting, the parameter is chosen as $|t'/t|=0.3$; i.e., $J^{[1]}_{\rm so}=0.6J$, and $J^{[2]}_{\rm so}=0.09J$. For a DQD with an interdot distance $2d=50$ nm, our theory gives $x_{\rm so}\approx180$ nm. This spin-orbit length is in good agreement with the experimental result $x_{\rm so}=230\pm50$ nm in Ref.~\cite{Nadj2}.

In the strong SOC regime with $|t'/t|=1$, $J=J^{[2]}_{\rm so}=(1/2)J^{[1]}_{\rm so}$ in Eq.~(\ref{EQ_anisotropic_exchange}). The effective Hamiltonian reads
\begin{equation}
H_{\rm eff}=E_{\rm qu}(S^{z}_{l}+S^{z}_{r})+J^{[1]}_{\rm so}(\textbf{S}_{l}\times\textbf{S}_{r})_{x}+J^{[1]}_{\rm so}S^{x}_{l}S^{x}_{r}.
\end{equation}
The energy spectrum in this case is shown in Fig.~\ref{Fig_anticrossing}(c). As in Figs.~\ref{Fig_anticrossing}(a) and~\ref{Fig_anticrossing}(b), the triplet state $T_{0}$ remains uncoupled to the singlet state $S$ and the triplet states $T_{\pm}$, but the other three eigenstates become superpositions of $S$ and $T_{\pm}$. Also, the level splitting at zero magnetic field changes from $J$ in Fig.~\ref{Fig_anticrossing}(a) to $J^{[1]}_{\rm so}$ in Fig.~\ref{Fig_anticrossing}(c). In this regime, the spectrum is similar to that in the weak SOC regime, but the anisotropic term $J^{[1]}_{\rm so}(\textbf{S}_{l}\times\textbf{S}_{r})_{x}$ now dominates in the exchange coupling.

In the ultrastrong SOC regime with $|t'/t|\gg1$, $J \ll\,J^{[1]}_{\rm so}\ll\,J^{[2]}_{\rm so}$ in Eq.~(\ref{EQ_anisotropic_exchange}). After neglecting the second-order terms, the effective Hamiltonian (\ref{EQ_anisotropic_exchange}) is reduced to
\begin{eqnarray}
H_{\rm eff}&=&E_{\rm qu}(S^{z}_{l}+S^{z}_{r})-J^{[2]}_{\rm so}\textbf{S}_{l}\cdot\textbf{S}_{r}+2J^{[2]}_{\rm so}S^{x}_{l}S^{x}_{r}\nonumber\\
&&+J^{[1]}_{\rm so}(\textbf{S}_{l}\times\textbf{S}_{r})_{x}.
\end{eqnarray}
The energy spectrum is shown in Fig.~\ref{Fig_anticrossing}(d). An apparent difference from the weak and strong SOC regimes is that both the singlet and triplet states, $S$ and $T_{0}$, become degenerate. Also, the zero-field level splitting is changed to $J^{[2]}_{\rm so}$. Currently, this ultrastrong SOC regime is unavailable in a semiconductor nanowire DQD, but it might be achievable in the future via quantum simulation in, e.g., ultracold-atom systems~\cite{Galitski}.

\section{\label{secIV}Conclusion}
We have studied the electron tunneling in a semiconductor nanowire DQD with strong SOC. In addition to the usual spin-conserved tunneling, there is also appreciable spin-flipped tunneling. While the total tunneling is irrelevant to the SOC, both the spin-conserved and spin-flipped tunnelings are SOC dependent and can compete with each other in the strong SOC regime. When two electrons are confined in this DQD, the lowest two states of each dot can be used to achieve a spin-orbit qubit. Within this DQD, the Coulomb repulsion between electrons can combine with the SOC-dependent tunnelings to yield an anisotropic Heisenberg exchange coupling between the two spin-orbit qubits. We obtain an analytical expression for this anisotropic exchange coupling, which is valid in the strong and even ultrastrong SOC regimes. Each exchange-coupling term has an explicit physical picture involving the second-order virtual tunneling, and its role varies in different SOC regimes. Our theory unveils some distinct properties of the nanowire DQD beyond the weak SOC regime.

\section*{Acknowledgements}
R.L. and J.Q.Y. are supported by National Natural Science Foundation of China Grant No.~91121015, National Basic Research Program of China Grant No.~2014CB921401, and NSAF Grant No.~U1330201.

\appendix

\section{\label{appendix_A}The orthonormal spin-orbit basis}

In this appendix, we orthogonalize the four states given in Eq.~(\ref{EQ_unorthonormal}) via the Schmidt orthogonalization method. Note that these four states are not orthogonal due to the overlap integrations $s_{a}$ and $s_{b}$ given in Eq.~(\ref{Eq_overlapintegration}).

For the states $\phi_{l\Downarrow}(x)$ and $\phi_{r\Downarrow}(x)$, using the conventional orthogonalization method~\cite{Burkard}, we obtain the following orthogonal states:
\begin{eqnarray}
\phi^{\rm or}_{l\Downarrow}(x)&=&\frac{1}{\sqrt{\zeta}}\big[\phi_{l\Downarrow}(x)-g_{a}\phi_{r\Downarrow}(x)\big],\nonumber\\
\phi^{\rm or}_{r\Downarrow}(x)&=&\frac{1}{\sqrt{\zeta}}\big[\phi_{r\Downarrow}(x)-g^{*}_{a}\phi_{l\Downarrow}(x)\big],
\end{eqnarray}
where
\begin{equation}
\zeta=1-2\mathrm{Re}(s_{a}g_{a})+|g_{a}|^{2},
\end{equation}
with $g_{a}=(1-\sqrt{1-s^{2}_{a}})/s_{a}$. Because of the overlap integration $s_{b}$, both $\phi_{l\Uparrow}(x)$ and $\phi_{r\Uparrow}(x)$ are not orthogonal to the states $\phi^{\rm or}_{l\Downarrow}(x)$ and $\phi^{\rm or}_{r\Downarrow}(x)$.

Our first step is to construct, via Schmidt orthogonalization, two intermediate states $\tilde{\phi}_{l\Uparrow}(x)$ and $\tilde{\phi}_{r\Uparrow}(x)$ which are orthogonal to the states $\phi^{\rm or}_{l\Downarrow}(x)$ and $\phi^{\rm or}_{r\Downarrow}(x)$; i.e.,
\begin{eqnarray}
\tilde{\phi}_{l\Uparrow}(x)&=&\frac{1}{\sqrt{1-\chi}}\left(\phi_{l\Uparrow}(x)
-\frac{s_{b}}{\sqrt{\zeta}}\left[\phi^{\rm or}_{r\Downarrow}(x)-g^{*}_{a}\phi^{\rm or}_{l\Downarrow}(x)\right]\right),\nonumber\\
\tilde{\phi}_{r\Uparrow}(x)&=&\frac{1}{\sqrt{1-\chi}}\left(\phi_{r\Uparrow}(x)-\frac{s_{b}}{\sqrt{\zeta}}
\left[\phi^{\rm or}_{l\Downarrow}(x)-g_{a}\phi^{\rm or}_{r\Downarrow}(x)\right]\right),\nonumber\\
\end{eqnarray}
where
\begin{equation}
\chi=\frac{(|g_{a}|^{2}+1)|s_{b}|^{2}}{1-2\mathrm{Re}(s_{a}g_{a})+|g_{a}|^{2}}~.
\end{equation}

Next, we orthogonalize these two states. It is easy to obtain the following orthogonal states:
\begin{eqnarray}
\phi^{\rm or}_{l\Uparrow}(x)&=&\frac{1}{\sqrt{\zeta'}}\left[\tilde{\phi}_{l\Uparrow}(x)-g'_{a}\tilde{\phi}_{r\Uparrow}(x)\right],\nonumber\\
\phi^{\rm or}_{r\Uparrow}(x)&=&\frac{1}{\sqrt{\zeta'}}\left[\tilde{\phi}_{r\Uparrow}(x)-g'^{*}_{a}\tilde{\phi}_{l\Uparrow}(x)\right],\nonumber\\
\end{eqnarray}
where
\begin{equation}
\zeta'=1-2\mathrm{Re}(s'_{a}g'_{a})+|g'_{a}|^{2},
\end{equation}
with $g'_{a}=(1-\sqrt{1-s'^{2}_{a}})/s'_{a}$, and
\begin{eqnarray}
s'_{a}&=&\int\,dx\tilde{\phi}^{\dagger}_{r\Uparrow}(x)\tilde{\phi}_{l\Uparrow}(x)\nonumber\\
&=&\frac{1}{1-\chi}\left(s_{a}+\frac{2g_{a}s^{*}_{b}s_{b}}{1-2\mathrm{Re}(s_{a}g_{a})+|g_{a}|^{2}}\right).
\end{eqnarray}
Obviously, $s'_{a}$ is the overlap integration between $\tilde{\phi}_{l\Uparrow}(x)$ and $\tilde{\phi}_{r\Uparrow}(x)$. Therefore, we have derived an orthonormal spin-orbit basis $\phi^{\rm or}_{k\sigma}(x)$, where $k=l,r$ and $\sigma=$~$\Uparrow,\Downarrow$.

\section{\label{appendix_B}The effective Hamiltonian}
Below we give the details for deriving the effective Hamiltonian $H_{\rm eff}$ from the second quantized Hamiltonian of a nanowire DQD. The DQD Hamiltonian can be written as $H=H_{s}+H_{t}+H_{U}$, with
\begin{eqnarray}
H_{s}&=&\sum_{\sigma}(\varepsilon_{l\sigma}c^{\dagger}_{l\sigma}c_{l\sigma}+\varepsilon_{r\sigma}c^{\dagger}_{r\sigma}c_{r\sigma}),\nonumber\\
H_{t}&=&\sum_{\sigma}(t_{\sigma}c^{\dagger}_{l\sigma}c_{r\sigma}+t'_{\sigma}c^{\dagger}_{l\sigma}c_{r\bar{\sigma}}+{\rm h.c.}),\\
H_{U}&=&Un_{l\Uparrow}n_{l\Downarrow}+Un_{r\Uparrow}n_{r\Downarrow}+U'\sum_{\sigma\sigma'}n_{l\sigma}n_{r\sigma'},\nonumber
\end{eqnarray}
where $\sigma=$~$\Uparrow$ and $\Downarrow$.

When the Coulomb repulsion in the DQD is so strong that $(U-U')\gg\,|t|,|t'|$, the two electrons in the DQD have a fixed charge configuration; i.e., each dot confines one and only one electron. Thus, we can define a projection operator~\cite{RuiLi2}
\begin{eqnarray}
P&=&[n_{l\Uparrow}(1-n_{l\Downarrow})+n_{l\Downarrow}(1-n_{l\Uparrow})]\nonumber\\
&&\times[n_{r\Uparrow}(1-n_{r\Downarrow})+n_{r\Downarrow}(1-n_{r\Uparrow})],
\end{eqnarray}
which retains the pseudospin degrees of freedom of the two electrons but reduces the Hilbert space to the subspace with each dot occupied by one electron.

The effective Hamiltonian can be written as~\cite{Nagaosa}
\begin{equation}
H_{\rm eff}=PHP-PHQ(QHQ-E)^{-1}QHP,
\end{equation}
where $Q=1-P$ and $E$ is the ground-state energy. The operator $QHQ-E$ describes the energy difference between the double- and single-electron occupations of the DQD, so $QHQ-E\approx\,U-U'$. Also, it is easy to know that $PHQ=PH_{t}Q$ and $QHP=QH_{t}P$. Therefore, we have
\begin{equation}
H_{\rm eff}=PHP-\frac{PH^{2}_{t}P}{U-U'}.
\end{equation}
After some algebra, we obtain
\begin{eqnarray}
PHP&=&PH_{s}P=(\varepsilon_{l\Uparrow}-\varepsilon_{l\Downarrow})S^{z}_{l}+(\varepsilon_{r\Uparrow}-\varepsilon_{r\Downarrow})S^{z}_{r}\nonumber\\
&&+\frac{\varepsilon_{l\Uparrow}+\varepsilon_{l\Downarrow}}{2}+\frac{\varepsilon_{r\Uparrow}+\varepsilon_{r\Downarrow}}{2},
\end{eqnarray}
and
\begin{figure}
\includegraphics{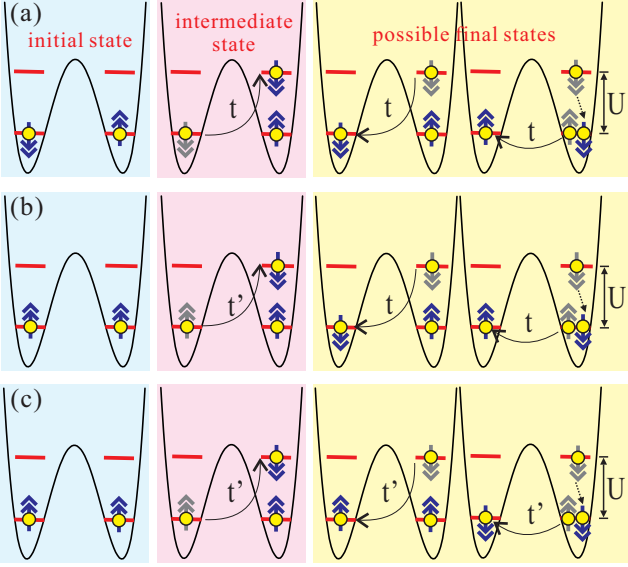}
\caption{\label{Fig_exchange_coupling}Schematical representation of the second-order virtual tunnelings involved in the exchange coupling between the two spin-orbit qubits. The virtual processes in (a) introduce an antiferromagnetic exchange coupling $J~\textbf{S}_{l}\cdot\textbf{S}_{r}$, the virtual processes in (b) introduce an anisotropic exchange coupling $J^{[1]}_{\rm so}(S^{y}_{l}S^{z}_{r}-S^{z}_{l}S^{y}_{r})$, and the virtual processes in (c) introduce a ferromagnetic exchange coupling $-J^{[2]}_{\rm so}\textbf{S}_{l}\cdot\textbf{S}_{r}+2J^{[2]}_{\rm so}S^{x}_{l}S^{x}_{r}$.}
\end{figure}
\begin{widetext}
\begin{eqnarray}
PH^{2}_{t}P&=&\sum_{\sigma,\sigma'}\big[|t|^{2}Pc^{\dagger}_{l\sigma}c_{l\sigma'}PPc_{r\sigma}c^{\dagger}_{r\sigma'}P+|t|^{2}
Pc^{\dagger}_{r\sigma}c_{r\sigma'}PPc_{l\sigma}c^{\dagger}_{l\sigma'}P+tt'^{*}Pc^{\dagger}_{l\sigma}c_{l\sigma'}PPc_{r\sigma}c^{\dagger}_{r\bar{\sigma'}}P\nonumber\\
&&+t^{*}t'Pc^{\dagger}_{r\sigma}c_{r\bar{\sigma'}}PPc_{l\sigma}c^{\dagger}_{l\sigma'}P+t't^{*}Pc^{\dagger}_{l\sigma}c_{l\sigma'}PPc_{r\bar{\sigma}}c^{\dagger}_{r\sigma'}P
+t'^{*}tPc^{\dagger}_{r\bar{\sigma}}c_{r\sigma'}PPc_{l\sigma}c^{\dagger}_{l\sigma'}P\nonumber\\
&&+|t'|^{2}Pc^{\dagger}_{l\sigma}c_{l\sigma'}PPc_{r\bar{\sigma}}c^{\dagger}_{r\bar{\sigma'}}P+|t'|^{2}Pc^{\dagger}_{r\bar{\sigma}}c_{r\bar{\sigma'}}PPc_{l\sigma}c^{\dagger}_{l\sigma'}P\big]
\nonumber\\
&=&\sum_{\sigma,\sigma'}\big[|t|^{2}(\delta_{\sigma\sigma'}/2+\textbf{S}_{l}\cdot\boldsymbol{\sigma}_{\sigma'\sigma})(\delta_{\sigma\sigma'}/2-\textbf{S}_{r}
\cdot\boldsymbol{\sigma}_{\sigma\sigma'})+
|t|^{2}(\delta_{\sigma\sigma'}/2+\textbf{S}_{r}\cdot\boldsymbol{\sigma}_{\sigma'\sigma})(\delta_{\sigma\sigma'}/2-\textbf{S}_{l}\cdot\boldsymbol{\sigma}_{\sigma\sigma'})\nonumber\\
&&+tt'^{*}(\delta_{\sigma\sigma'}/2+\textbf{S}_{l}\cdot\boldsymbol{\sigma}_{\sigma'\sigma})(\delta_{\sigma\bar{\sigma'}}/2-\textbf{S}_{r}\cdot
\boldsymbol{\sigma}_{\sigma\bar{\sigma'}})+t^{*}t'(\delta_{\sigma\bar{\sigma'}}/2+\textbf{S}_{r}\cdot\boldsymbol{\sigma}_{\bar{\sigma'}\sigma})
(\delta_{\sigma\sigma'}/2-\textbf{S}_{l}\cdot\boldsymbol{\sigma}_{\sigma\sigma'})\nonumber\\
&&+t't^{*}(\delta_{\sigma\sigma'}/2+\textbf{S}_{l}\cdot\boldsymbol{\sigma}_{\sigma'\sigma})(\delta_{\bar{\sigma}\sigma'}/2-\textbf{S}_{r}\cdot
\boldsymbol{\sigma}_{\bar{\sigma}\sigma'})+t'^{*}t(\delta_{\bar{\sigma}\sigma'}/2+\textbf{S}_{r}\cdot\boldsymbol{\sigma}_{\sigma'\bar{\sigma}})
(\delta_{\sigma\sigma'}/2-\textbf{S}_{l}\cdot\boldsymbol{\sigma}_{\sigma\sigma'})\nonumber\\
&&+|t'|^{2}(\delta_{\sigma\sigma'}/2+\textbf{S}_{l}\cdot\boldsymbol{\sigma}_{\sigma'\sigma})(\delta_{\bar{\sigma}\bar{\sigma'}}/2-\textbf{S}_{r}
\cdot\boldsymbol{\sigma}_{\bar{\sigma}\bar{\sigma'}})+|t'|^{2}(\delta_{\bar{\sigma}\bar{\sigma'}}/2+\textbf{S}_{r}\cdot\boldsymbol{\sigma}_{\bar{\sigma'}
\bar{\sigma}})(\delta_{\sigma\sigma'}/2-\textbf{S}_{l}\cdot\boldsymbol{\sigma}_{\sigma\sigma'})\big]\nonumber\\
&=&2|t|^{2}(\frac{1}{2}-2\textbf{S}_{l}\cdot\textbf{S}_{r})+2|t'|^{2}(\frac{1}{2}+2\textbf{S}_{l}\cdot\textbf{S}_{r}-4S^{x}_{l}
S^{x}_{r})+(4itt'^{*}-4it^{*}t')(S^{z}_{l}S^{y}_{r}-S^{y}_{l}S^{z}_{r}),
\end{eqnarray}
\end{widetext}
where $\textbf{S}_{k=l,r}=(1/2)\sum_{\sigma,\sigma'}c^{\dagger}_{k\sigma}\boldsymbol{\rho}_{\sigma\sigma'}c_{k\sigma'}$ is the pseudospin operator, with $\boldsymbol{\rho}=(\rho^{x},\rho^{y},\rho^{z})$ being the Pauli matrices of the spin-orbit qubit.
Thus, we have the following effective Hamiltonian describing the pseudospin degrees of freedom of the two electrons confined in the DQD:
\begin{eqnarray}
H_{\rm eff}&=&E_{\rm qu}(S^{z}_{l}+S^{z}_{r})+(J-J^{[2]}_{\rm so})\textbf{S}_{l}\cdot\textbf{S}_{r}\nonumber\\
&&+J^{[1]}_{\rm so}(\textbf{S}_{l}\times\textbf{S}_{r})_{x}+2J^{[2]}_{\rm so}S^{x}_{l}S^{x}_{r}.\label{EQ_anisotropic_exchange_ap}
\end{eqnarray}
This is the effective Hamiltonian $H_{\rm eff}$ in Eq.~(\ref{EQ_anisotropic_exchange}), with $E_{\rm qu}$, $J$, $J^{[1]}_{\rm so}$, and $J^{[2]}_{\rm so}$ given in Eq.~(\ref{Eq_coefficients}).

Each exchange-coupling term in Eq.~(\ref{EQ_anisotropic_exchange_ap}) is induced by the second-order virtual tunnelings in a DQD. The virtual tunneling involving $t^{2}$ gives an antiferromagnetic exchange interaction $J~\textbf{S}_{l}\cdot\textbf{S}_{r}$ [see Figs.~\ref{Fig_exchange_coupling}(a)], the virtual tunneling involving the combination of $t$ and $t'$ gives an anisotropic exchange interaction $J^{[1]}_{\rm so}(\textbf{S}_{l}\times\textbf{S}_{r})_{x}$ [see Fig.~\ref{Fig_exchange_coupling}(b)], and the virtual tunneling involving $t'^{2}$ gives a ferromagnetic exchange interaction $-J^{[2]}_{\rm so}\textbf{S}_{l}\cdot\textbf{S}_{r}+2J^{[2]}_{\rm so}S^{x}_{l}S^{x}_{r}$ [see Fig.~\ref{Fig_exchange_coupling}(c)]. Note that in the absence of SOC, $t'=0$, so $J^{[1]}_{\rm so}=J^{[2]}_{\rm so}=0$ [see Eq.~(\ref{Eq_coefficients})]. Therefore, without the SOC, only the isotropic antiferromagnetic term $J~\textbf{S}_{l}\cdot\textbf{S}_{r}$ occurs in the exchange coupling. Here we take Fig.~\ref{Fig_exchange_coupling}(b) as an example. Given an initial two-qubit state $|\Uparrow_{l}\Uparrow_{r}\rangle$, under the virtual tunnelings, the possible final states are $|\Downarrow_{l}\Uparrow_{r}\rangle$ and $|\Uparrow_{l}\Downarrow_{r}\rangle$. This virtual tunneling can be described by the action of the operator $J^{[1]}_{\rm so}(\textbf{S}_{l}\times\textbf{S}_{r})_{x}$ on the state $|\Uparrow_{l}\Uparrow_{r}\rangle$:
\begin{equation}
J^{[1]}_{\rm so}(\textbf{S}_{l}\times\textbf{S}_{r})_{x}|\Uparrow_{l}\Uparrow_{r}\rangle=i\frac{J^{[1]}_{\rm so}}{4}\left(|\Downarrow_{l}\Uparrow_{r}\rangle-|\Uparrow_{l}\Downarrow_{r}\rangle\right).
\end{equation}
Similarly, Figs.~\ref{Fig_exchange_coupling}(a) and \ref{Fig_exchange_coupling}(c) can also be explained this way.

\end{document}